\documentclass{PoS}

\title{Top Quark Properties}

\ShortTitle{Top Quark Properties}

\author{\speaker{Andreas B.\,Meyer}\\
        {\rm On behalf of the ATLAS, CDF, CMS and D0 Collaborations} \\
        DESY\\
        E-mail: \email{andreas.meyer@cern.ch}}

\abstract{Recent measurements of top-quark properties at the LHC and at the Tevatron are presented. The results include precision measurements of standard model parameters, such as the top-quark mass, the measurement of angular distributions as well as the search for anomalous couplings.}

\FullConference{International Symposium on Lepton Photon Interactions at High Energies\\
		17-22 August 2015\\
		University of Ljubljana, Slovenia}

\begin{document}

\section{Introduction}

The top quark is by far the heaviest known elementary particle. Due to its large mass it is expected to couple strongly to the Higgs boson and to play a crucial role in electroweak loop corrections. The large top quark mass leads to a peculiar hierarchy for the timescales of production ($1/m_t$), decay ($1/\Gamma_t$), hadronisation ($1/\Lambda_{QCD}$) and spin decorrelation time ($m_t/\Lambda^2$) which in turn provides the top quark with unique features. Due to their short decay lifetime, top-quark events offer direct experimental access to properties of the top quark such as spin and charge. Not least, since the top quark is a coloured particle, top-quark measurements provide important input to QCD calculations. 

According to the standard model, top quarks decay almost exclusively to a W-boson and b-quark. The decay channel of the W-boson into leptons or quarks is then generally used to distinguish different top-quark decay channels. At the Tevatron, top quarks were produced in proton--anti-proton collisions at a centre-of-mass energies of 1.8 and 1.96 TeV, and the dominant contribution to the top-quark pair cross section is from $q\bar{q}$ annihilation. At the LHC, proton--proton collisions were measured at energies of 7 and 8 TeV, and since 2015 at 13 TeV. At these higher energies the top-quark pair cross section is significantly larger and dominantly driven by gluon-gluon fusion. Single top-quark production can occur via electro-weak interactions. Recent results on the production of top quarks are reported in another contribution~\cite{cristinziani}. Between 2010 and 2012 more than 5 million top-quark events were produced in proton--proton collisions by the ATLAS and CMS experiments each~\cite{atlas,cms}, exceeding the statistics of the complementary proton--anti-proton samples from the Tevatron by about two orders of magnitude.

Based on this wealth of data, top-quark physics has entered a new realm of precision, and experimental results are used to further constrain the standard model parameters, to probe improved QCD calculations and to search for new physics signals. Especially in scenarios in which new physics would couple to mass, the top quark would exhibit particular sensitivity. In this presentation, the most recent results of top-quark property measurements and searches for anomalous top-quark couplings from the Tevatron and the LHC are presented.


\section{Angular Distributions}

The measurement of angular distributions of top-quark final states are key to testing fundamental properties of the top quark in production and decay. In this section, recent measurements of spin correlations and asymmetries between top quarks and anti-quarks are reported.

\subsection{Spin Correlations}

\begin{figure}[tbp]
  \begin{center}
	\includegraphics[width=0.44\textwidth]{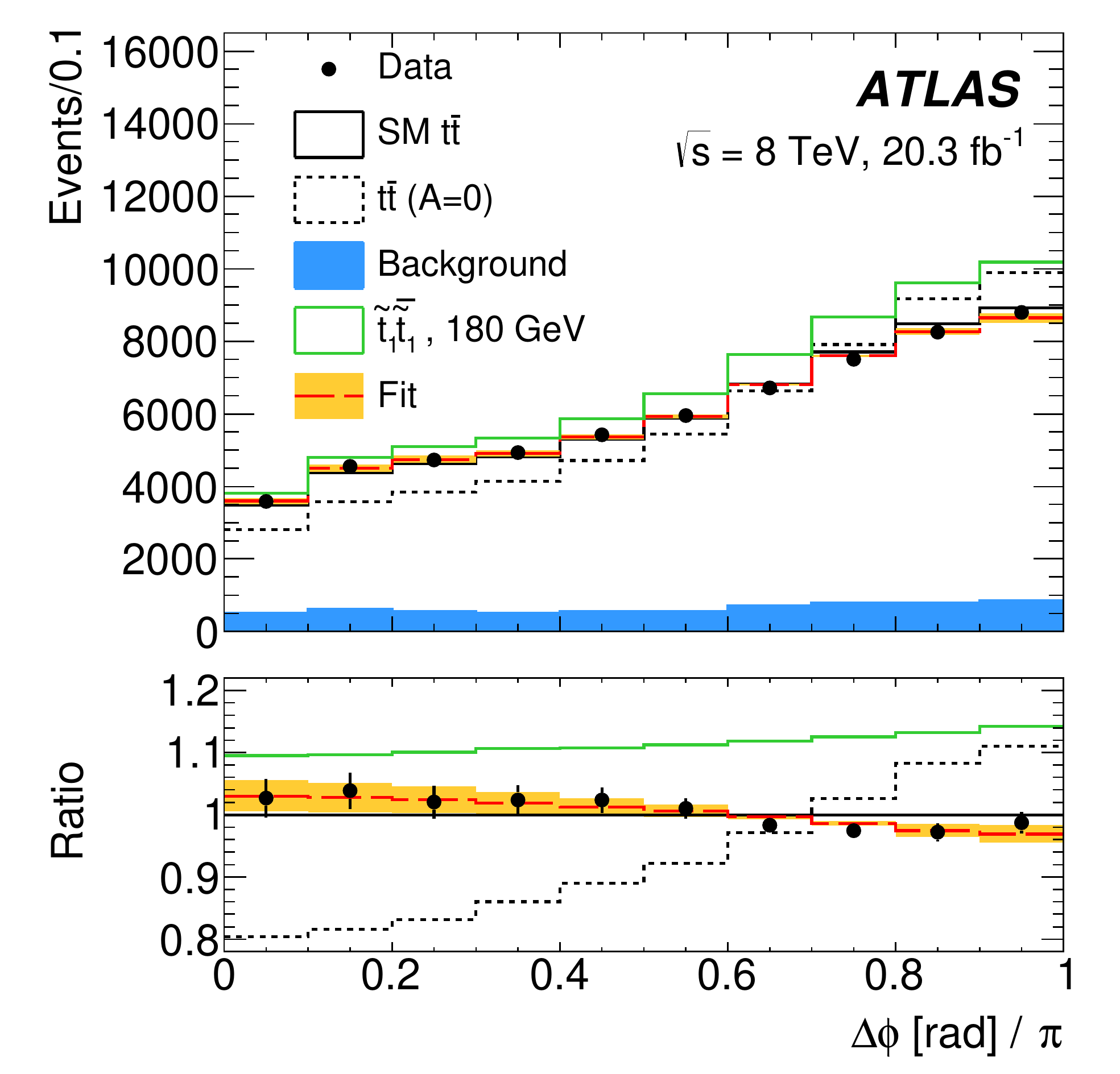}
	\includegraphics[width=0.55\textwidth]{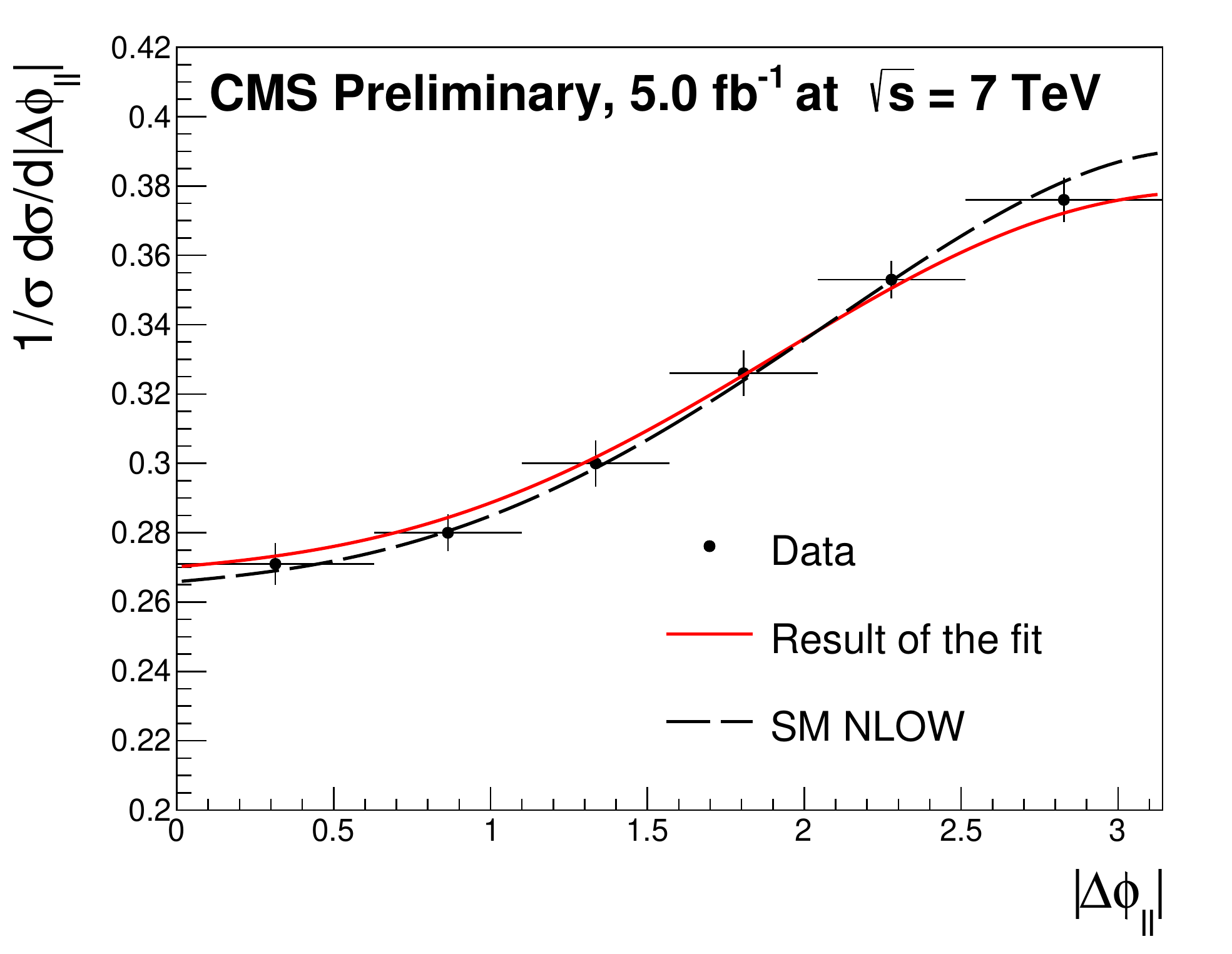}
	\caption{Distributions of the azimuthal angle between the two leptons in the laboratory system for (left) the ATLAS measurement~\cite{Aad:2014mfk} and (right) the CMS measurement~\cite{Chatrchyan:2013wua, cms-top-14-005}.}
    \label{fig:top:SC1}
  \end{center}
\end{figure}

If top quarks are spin-1/2 particles that behave according to standard model expectations, then they should be unpolarized in $t\bar{t}$ production, and their spins should be correlated. This expectation can be tested by measuring the angular decay distributions of the leptons from the decays of the W bosons which carry the information on the spin history of the event. In Figure~\ref{fig:top:SC1} (left), an ATLAS measurement of $t\bar{t}$ events with two leptons in the final state is presented~\cite{Aad:2014mfk}. The distribution of the azimuthal angle between the two leptons in the laboratory system is displayed together with the predictions from the standard model. Scenarios with no correlation and with contributions from stop quarks decaying to top quarks are also shown. Consistency with the standard model expectation can be quantified by a fit of the fraction of events that show SM spin correlations, $f_{SM}=N_{SM}/(N_{SM}+N_{\rm uncor})$, where $N_{SM}$ is the distribution of $t\bar{t}$ events as expected in the standard model and $N_{\rm uncor}$ is the distribution of uncorrelated $t\bar{t}$ events. The ATLAS experiment measures $f_{SM}$ to be $ 1.20 \pm 0.05(stat) \pm 0.13(syst)$. Contributions from possible supersymmetric partners of the top quark, the stops, would produce an uncorrelated component. Assuming the minimally supersymmetric standard model, MSSM, for the production of stops, and a stop-to-top branching ratio of 100\%, stop contributions can be excluded at the 95\% confidence level for stop masses between the top-quark mass and 191 GeV.

CMS performed a similar measurement of spin correlations using data at 7 TeV~\cite{Chatrchyan:2013wua}. The corresponding distribution is shown in Figure~\ref{fig:top:SC1} (right). The fraction $f_{SM}$ is determined to be $1.02 \pm 0.10 \pm 0.22$. CMS uses this experimental result to set a limit on new physics in the form of a top quark chromo-magnetic anomalous coupling. The limit on the real part $\Re$ of the chromo-magnetic dipole moment $\mu_t$ is determined to be $−0.043 < \Re(\mu_t) < 0.117$ at 95\% C.L.~\cite{cms-top-14-005}.
\begin{figure}[tbp]
  \begin{center}
	\includegraphics[width=0.55\textwidth]{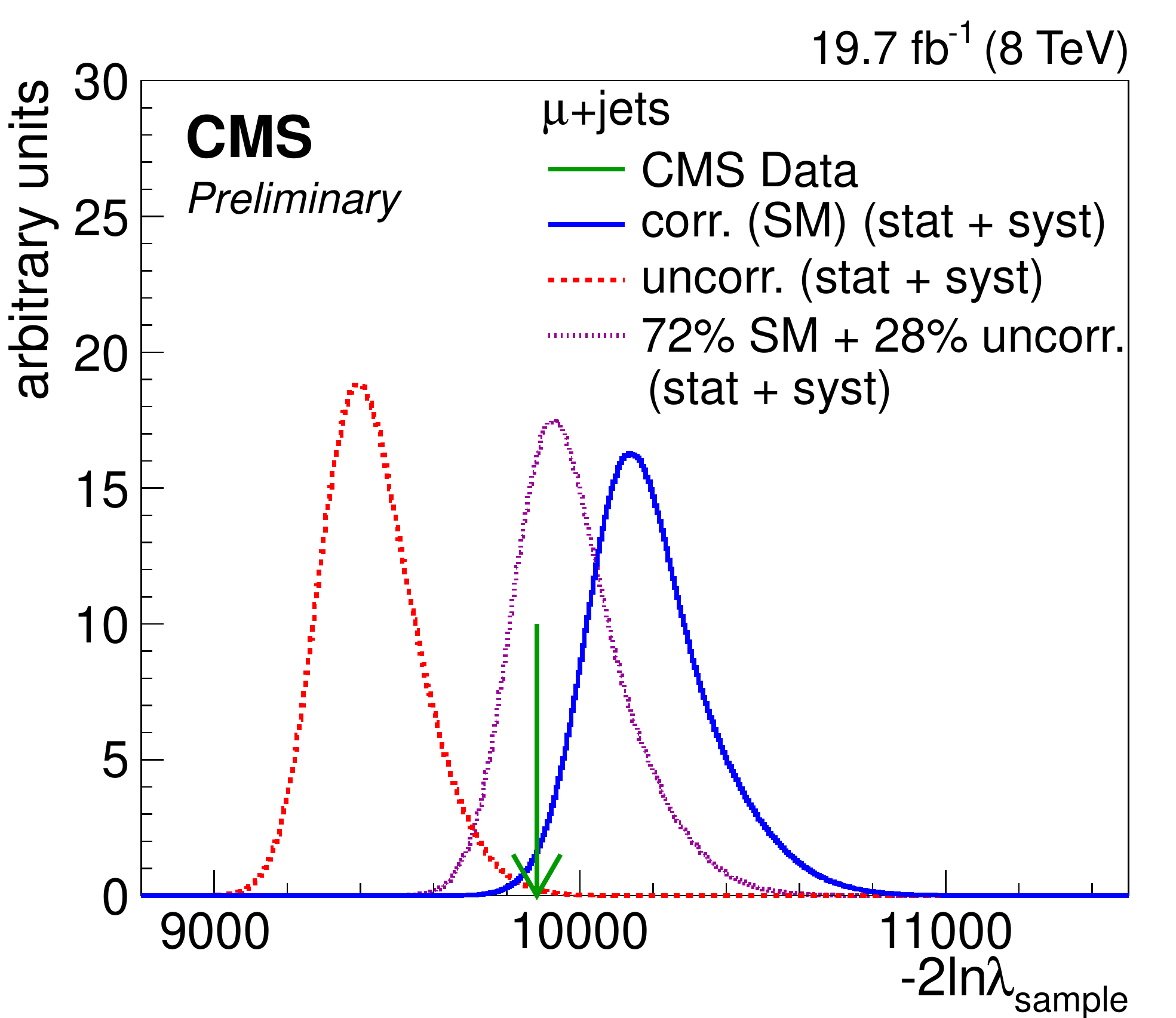}
	\includegraphics[width=0.4\textwidth]{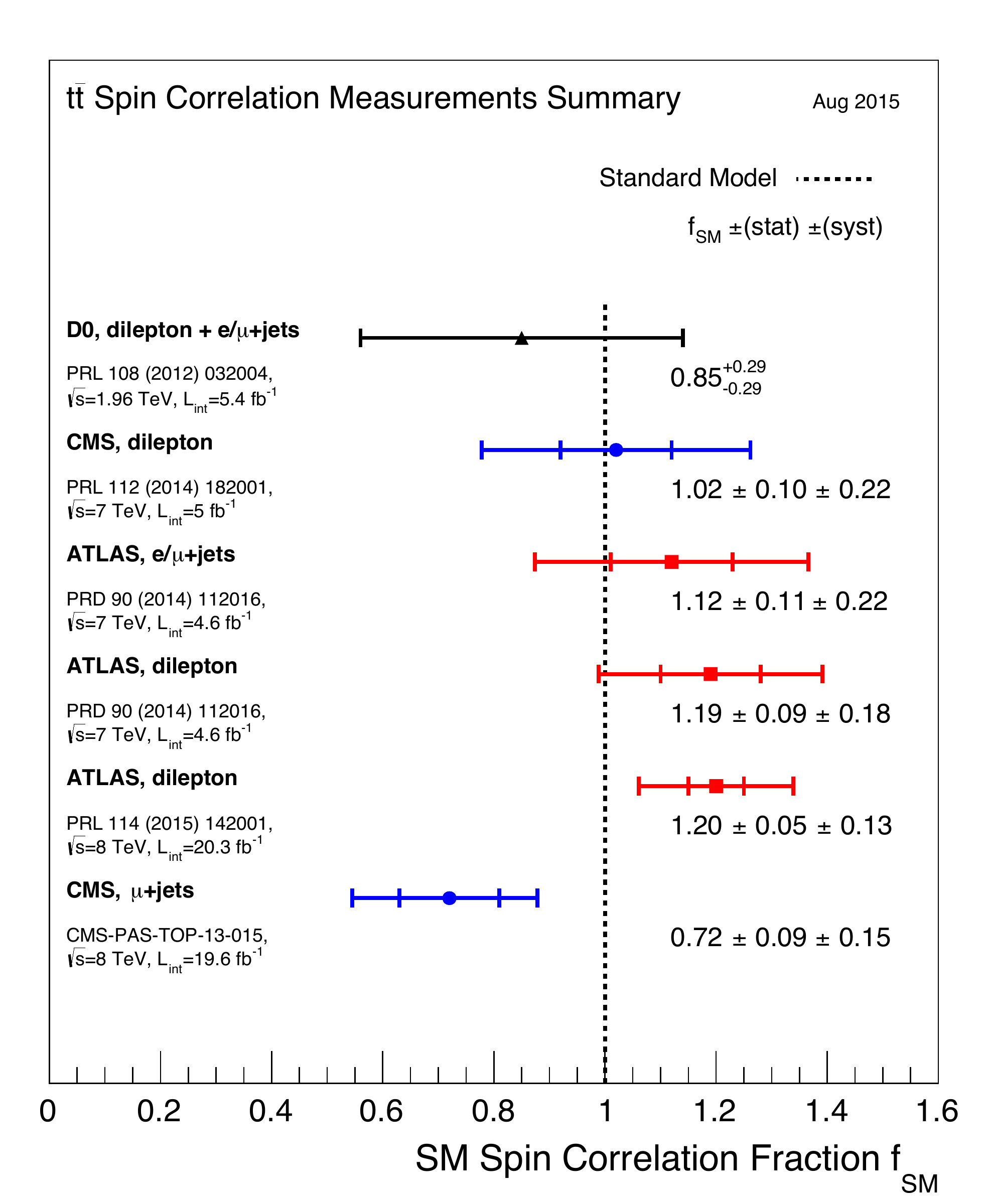}
	\caption{Left: sample hypothesis likelihood distributions, including the statistical and systematic uncertainties of the measurement. The solid (dotted) lines refer to the standard model (uncorrelated) hypothesis distributions, respectively. The result of the measurement is indicated by the arrow. Also shown is the expected distribution assuming $f_{SM}=0.72$~\cite{cms-top-13-015}. Right: summary of measurements of $f_{SM}$ at the Tevatron and the LHC.}
  \label{fig:top:SC2}
  \end{center}
\end{figure}

In a recent CMS analysis the consistency of the top-quark pair events produced with standard model (SM) correlated spins is tested in the $\mu$+jets final state~\cite{cms-top-13-015}. 
The measurement is obtained using a matrix element method and the result is presented in Figure~\ref{fig:top:SC2} (left). The data agree with the SM hypothesis within 2.2 standard deviations. This uncertainty includes statistical and systematic uncertainties. In addition, using a template fit method, the fraction of events which show SM spin correlations $f_{SM}$ is determined to be $0.72 \pm 0.09(stat)^{+0.15}_{-0.13}(syst)$. This result represents the most precise measurement of this quantity in the lepton+jets channel to-date. A summary of the spin correlation measurements performed so far at the Tevatron and the LHC is shown in Figure~\ref{fig:top:SC2} (right).

\subsection{Forward-Backward and Charge Asymmetry}

In the standard model, at leading order, top-quark pairs are produced in a symmetric state. The initial-state processes are $q\bar{q} \rightarrow t\bar{t}$, and gluon-fusion, $gg \rightarrow t\bar{t}$. At next-to-leading order, additional diagrams arise, and charge asymmetry is induced by interference between processes with initial-state and final-state radiation and between tree and box diagrams~\cite{acth}. Charge asymmetry results in an asymmetry of the $t\bar{t}$ event kinematics: top quarks (anti-quarks) are preferentially emitted in the direction of the incoming quark (anti-quark), respectively. At the Tevatron experiments, the initial state is $p\bar{p}$ and experimentally a forward-backward asymmetry can be measured as $N((\Delta y > 0) - N(\Delta y < 0))/N((\Delta y > 0) + N(\Delta y < 0))$, where $\Delta y$ is the difference between the signed rapidities of top quark and anti-quark and $N$ is the number of events. In contrast, at the LHC, the initial state is symmetric, and a charge asymmetry is induced only from the (momentum) difference of the (valence+sea) quark and (sea) anti-quark distributions in the proton, leading to a difference of absolute rapidities of top quarks and anti-quarks, $|y_t|-|y_{\bar{t}}|$.

\begin{figure}[tbp]
  \begin{center}
	\includegraphics[width=0.5\textwidth]{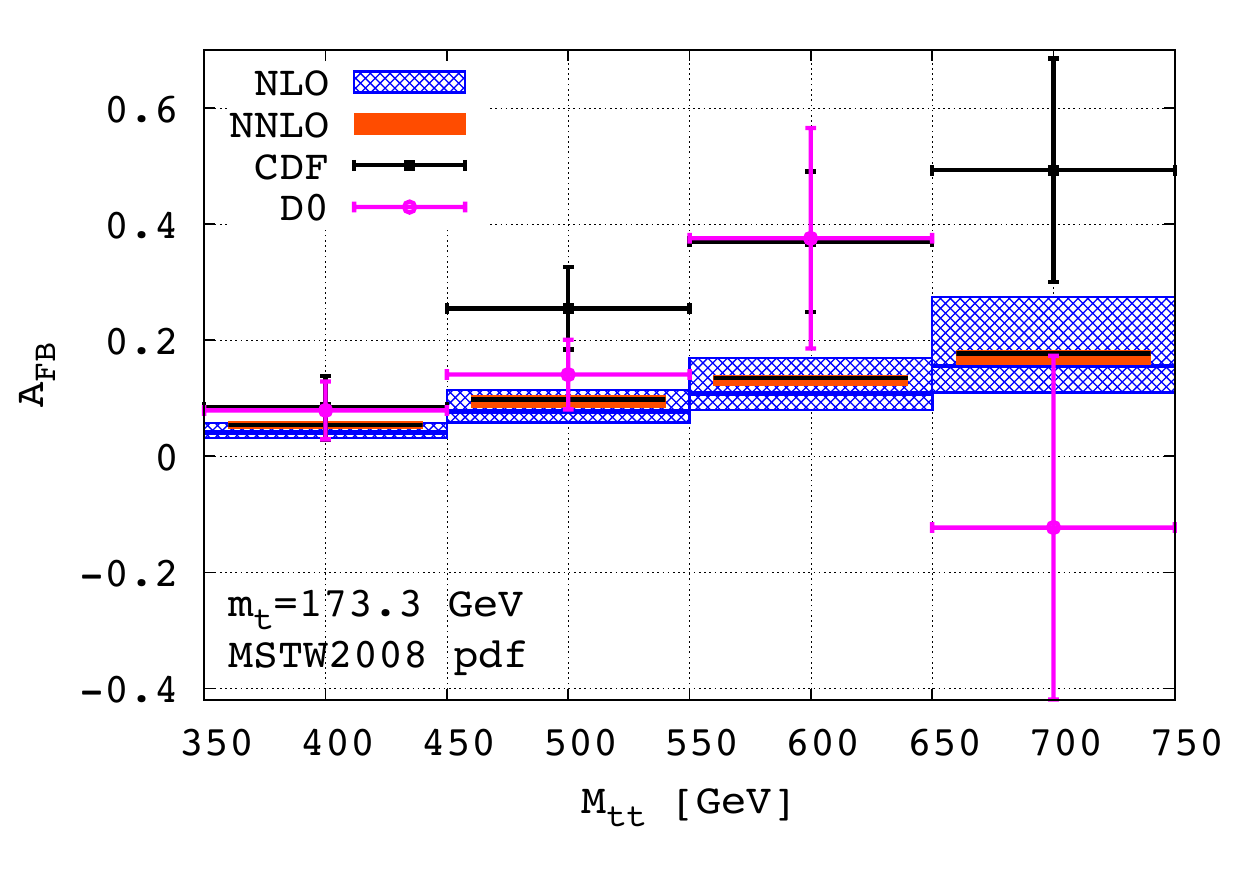}
	\includegraphics[width=0.42\textwidth]{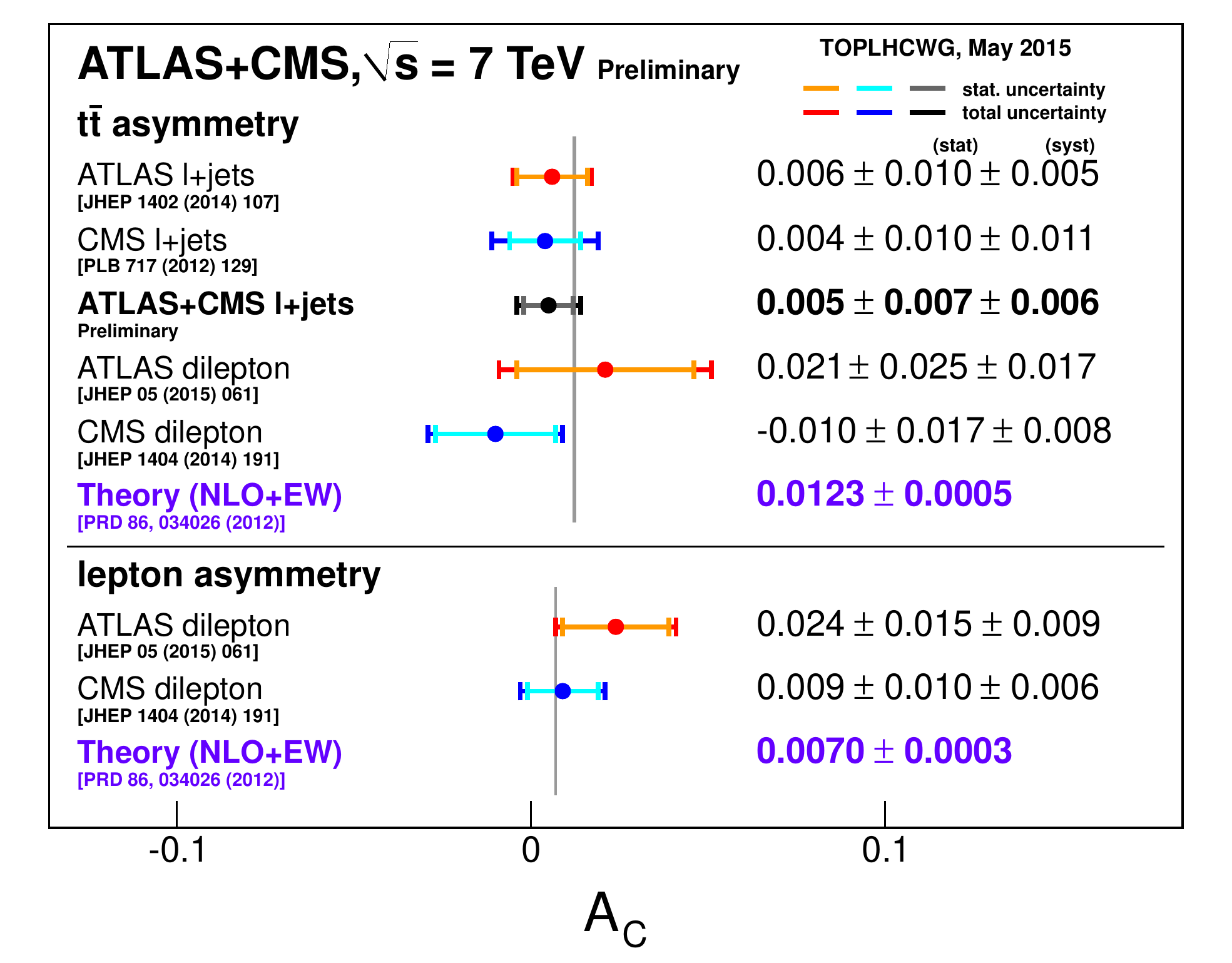}
	\caption{Left: forward-backward asymmetry $A_{\rm FB}$ as a function of the invariant mass of the $t\bar{t}$ system. The most recent Tevatron data (points) are compared with the NNLO prediction (taken from~\cite{Czakon:2014xsa}). Right: summary of measurements of $A_C$ at a centre-of-mass energy of 7 TeV at the LHC~\cite{lhctopwg}.}
    \label{fig:top:AC}
  \end{center}
\end{figure}

Most recently the predictions for the forward-backward asymmetry at the Tevatron have been calculated to next-to-next-to-leading order (NNLO). For the inclusive asymmetry $A_{\rm FB}$ a value of $0.095 \pm 0.007$ is expected~\cite{Czakon:2014xsa}. The distribution of $A_{\rm FB}$ as a function of the invariant mass of the top-quark pair is shown in Figure~\ref{fig:top:AC} (left), together with the most recent measurements from the CDF and D0 experiments~\cite{Aaltonen:2012it,Abazov:2014cca}. Data and theory are seen to agree within 1.5$\sigma$. Further details are described in another contribution~\cite{melnikov}. A summary of the results for the charge asymmetry $A_C$ from ATLAS and CMS experiments for a proton--proton center-of-mass energy of 7 TeV is given in Figure~\ref{fig:top:AC} (right). 

\begin{figure}[tbp]
  \begin{center}
	\includegraphics[width=0.4\textwidth]{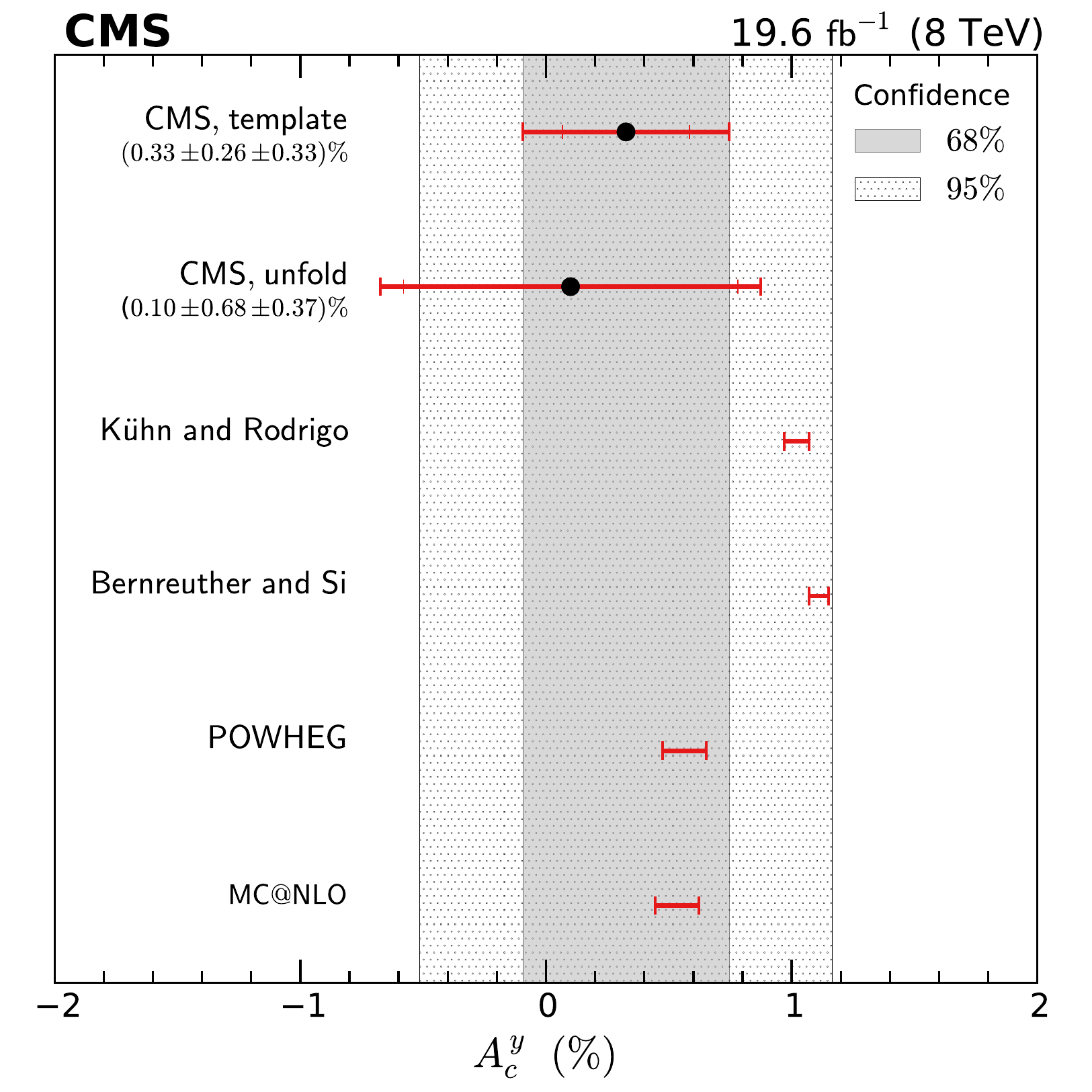}
	\includegraphics[width=0.56\textwidth]{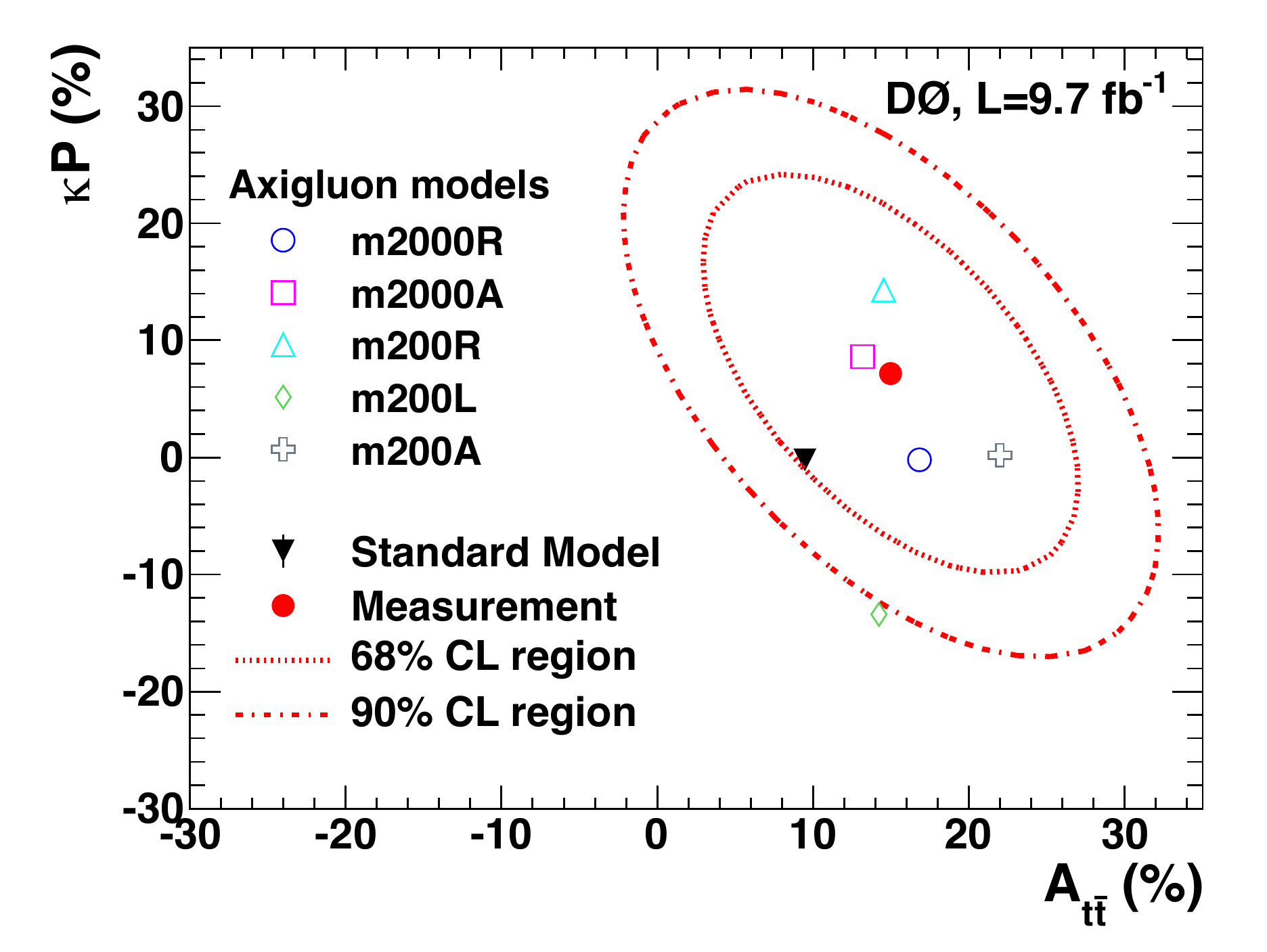}
	\caption{Left: measurements of the inclusive charge asymmetry at a centre-of-mass energy of 8 TeV by the CMS experiment~\cite{Khachatryan:2015mna}. The results from the template and unfolding approaches are compared with several predictions at next-to-leading order accuracy. Right: two-dimensional visualisation of the $A_{\rm FB}$ and top-quark polarization measurement and comparison with benchmark new physics models from the D0 experiment~\cite{Abazov:2015fna}.}
    \label{fig:top:pol}
  \end{center}
\end{figure}

New measurements of the charge asymmetry in the 8 TeV data are available from the CMS experiment, using two different approaches. An unfolding procedure is applied to correct for detector effects, and the charge asymmetry is measured both inclusively and differentially as a function of the kinematics of the $t\bar{t}$ system~\cite{Khachatryan:2015oga}. For the inclusive asymmetry, a significantly more precise result is obtained using a template-based approach in which the data are confronted with predictions at reconstruction level and the shape of the asymmetric component of the distribution is taken into account~\cite{Khachatryan:2015mna}. The results for the inclusive charge asymmetry are displayed in Figure~\ref{fig:top:pol} (left).

Most recently the D0 experiment presented a simultaneous measurement of the forward-backward asymmetry $A_{\rm FB}$ and the top-quark polarization in events with two leptons in the final state~\cite{Abazov:2015fna}. The analysis makes use of a matrix element technique to calculate likelihoods of the possible $t\bar{t}$ kinematic configurations. The result is presented in Figure~\ref{fig:top:pol} (right). It agrees within 1$\sigma$ with the standard model predictions. Assuming the top-quark polarization to be zero, as expected in the standard model, $A_{\rm FB}$ is measured to be $(17.5 \pm 5.6 (stat) \pm 3.1 (syst))\%$.

\section{Rare Decays}

New physics could manifest itself in rare decays of the top quark, i.e.\,in the enhancement of branching ratios for processes which in the standard model are expected to be vanishingly small.
Flavour-changing neutral currents (FCNC) are highly suppressed in the SM but very large enhancements could be realised according to many new physics models, and the LHC data are expected to be able to discover or exclude some of these models~\cite{AguilarSaavedra:2004wm}.

\begin{figure}[tbp]
  \begin{center}
	\includegraphics[width=0.52\textwidth]{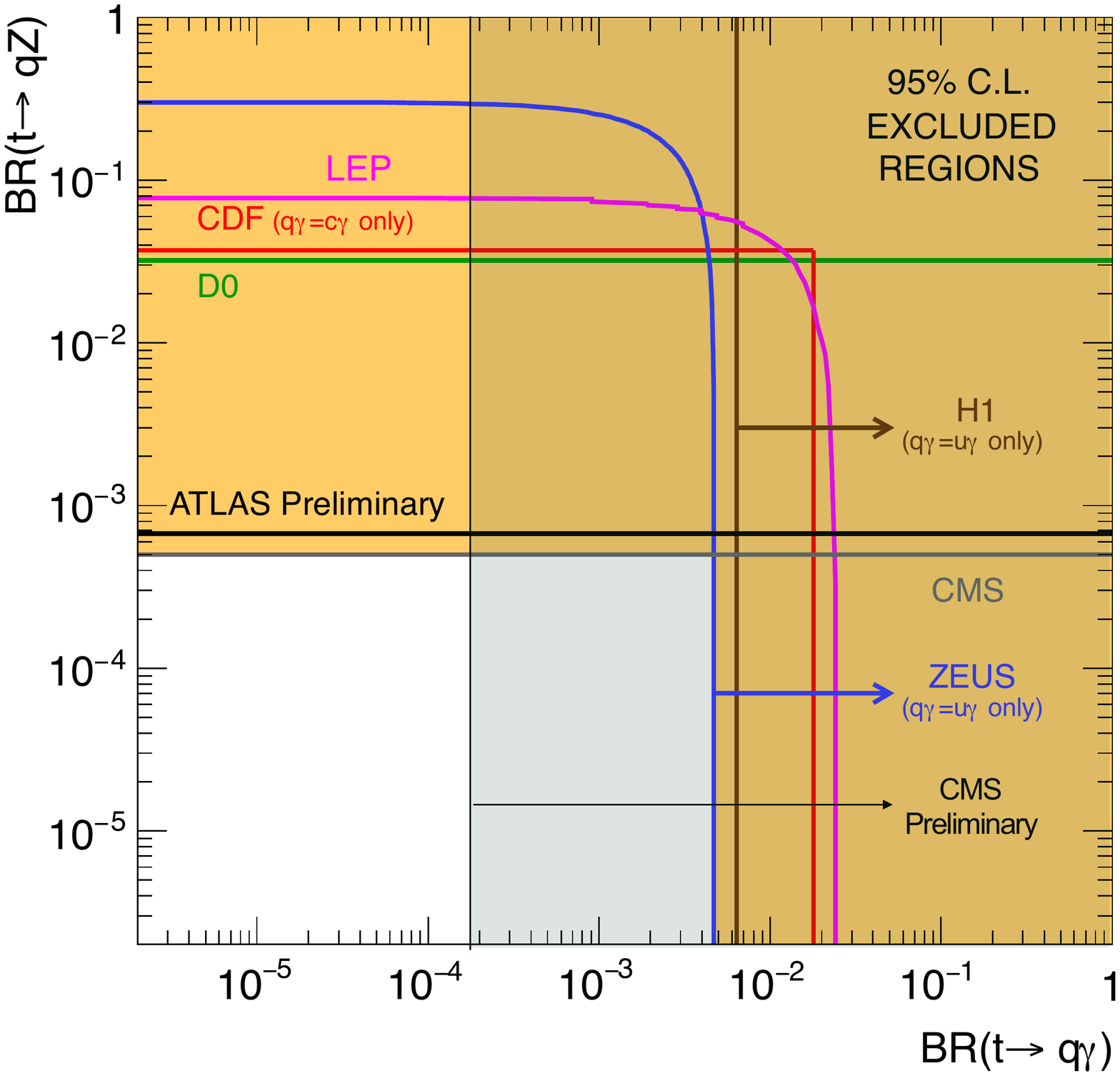}
	\includegraphics[width=0.47\textwidth]{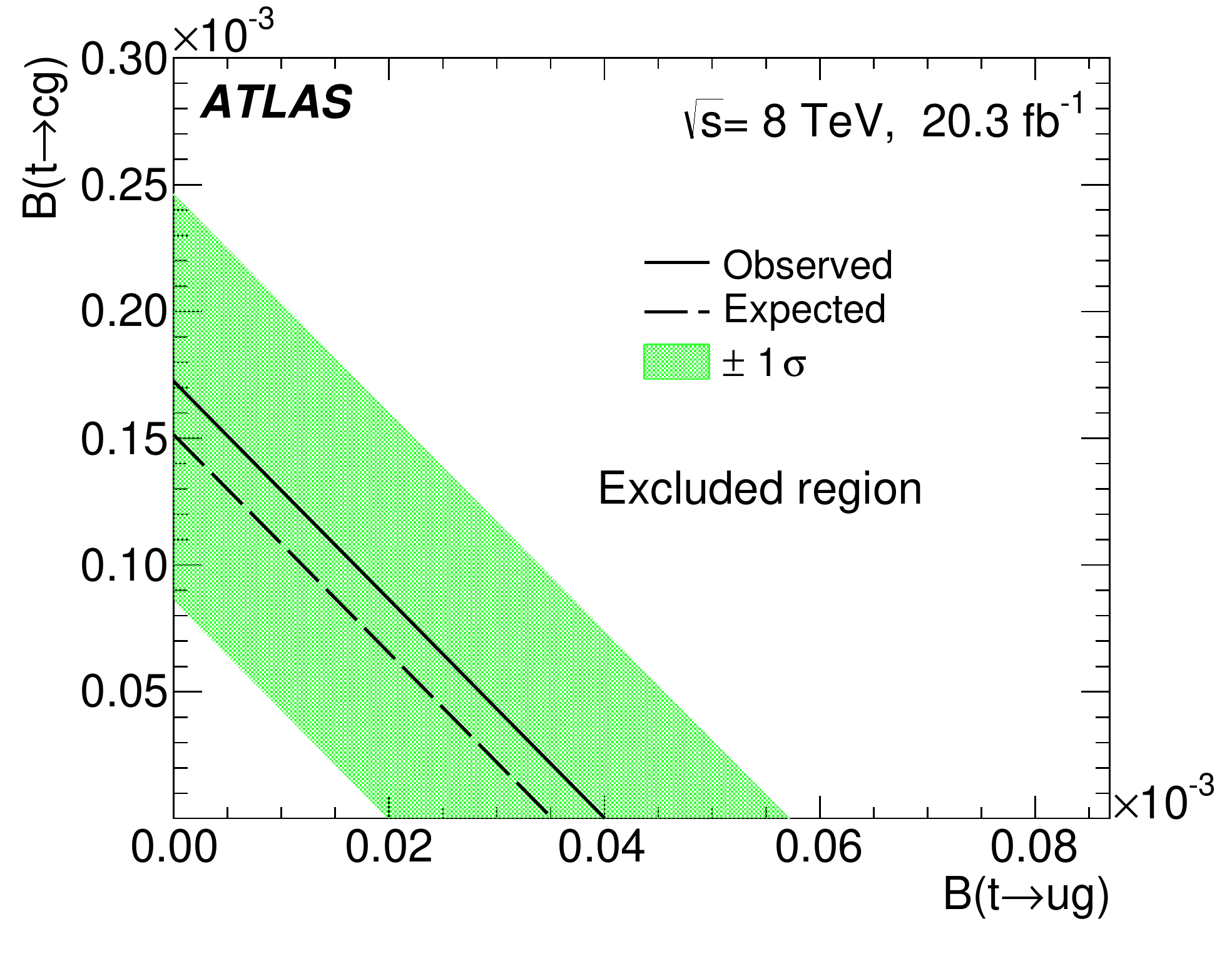}
	\caption{Left: summary of results from searches for flavour-changing neutral currents at HERA, LEP, Tevatron and LHC experiments. The original figure from~\cite{Aad:2015uza} has been modified to include the result from~\cite{cms-top-14-003}. For reference to the other results, please see references in~\cite{Aad:2015uza}. Right: upper limit on the branching fractions $BR(t \rightarrow ug)$ and
$ BR(t \rightarrow cg)$. The shaded band shows the one standard deviation variation of the expected limit~\cite{Aad:2015gea}.}
    \label{fig:top:fcnc}
  \end{center}
\end{figure}

ATLAS recently published a search for the flavour-changing neutral-current (FCNC) decay $t \rightarrow qZ$, where $q$ is an up or a charm quark, using the full 8 TeV dataset recorded in 2012~\cite{Aad:2015uza}. Top-quark pair-production events are used to search for final state topologies with three leptons where one lepton comes from the top-quark decay into $W$ boson and $b$ quark, as expected in the standard model, and two leptons arise from the decay of the $Z$ boson. Dominant standard model backgrounds arise from $WZ$, $ttV$ and $tZ$ production and their contribution is determined from control regions.
An observed (expected) limit on the branching ratio of 0.07\% (0.08\%) is set at 95\% C.L.. In Figure~\ref{fig:top:fcnc} (left) a summary of the results from searches for FCNC $t \rightarrow qZ$ and $t \rightarrow q\gamma$ is shown.

Flavour-changing neutral currents could occur not only in the decays of top quarks but also in their production, and events in which single top-quarks are produced are particularly suited for the search. The CMS experiment recently presented a search for top-quark production in association with a photon~\cite{cms-top-14-003}. The analysis makes use of a multi-variate analysis technique to separate signal from background. Observed (expected) limits on the branching ratio are determined, yielding $BR(t\rightarrow u\gamma) <0.0161\% (0.0279\%)$ and $BR(t\rightarrow u\gamma) <0.182\% (0.261\%)$, respectively. The result is also shown in Figure~\ref{fig:top:fcnc} (left).

Recent results are also available for the search of the process $t\rightarrow gq$. 
Both ATLAS and CMS perform the search in single top-quark production where the top quark would be produced in proton--proton collisions from the coupling of an initial state gluon with an up- or charm-quark~\cite{Aad:2015gea,cms-top-14-007}. Candidate events are classified into signal- and background-like candidates using neural network techniques. CMS so far reports results based on the 7 TeV dataset using decays of the top quark in the muon channel only. The ATLAS experiment performs the analysis on the full 8 TeV dataset using both electron and muon channels, and yields the most stringent limits on the branching fractions of $BR(t \rightarrow ug) < 4.0 \times 10^{-5}$ and $BR(t \rightarrow cg) < 17 \times 10^{-5}$. This result is also shown in Figure~\ref{fig:top:fcnc} (right).

Searches are also performed for scenarios where the top-quark decays to a quark (up or charm) and a neutral Higgs boson. Both ATLAS and CMS report searches for $t\bar{t}$ events in which one top-quark decays to $qH$ and the other decays to $bW$. The Higgs boson is identified through its decay into two photons~\cite{Aad:2014dya,cms-top-14-019}. In these analyses both the hadronic and the leptonic decay modes of the W boson are used. In the absence of a signal, observed (expected)  upper limits are determined on the branching ratios $BR(t \rightarrow qH) < 0.79\% (0.51\%)$ for ATLAS, and $BR(t\rightarrow cH) < 0.47\% (0.71\%)$ or $BR(t\rightarrow uH) < 0.42\% (0.65\%)$ for CMS.
Other decay channels of the Higgs into pairs of $Z$, $W$ or $\tau$-leptons, yielding multi-lepton final states have also been explored~\cite{cms-top-13-017}.

While none of the many searches for flavour-changing neutral currents has been successful yet, the achieved upper limits are starting to be close to expectations for some of the new physics scenarios~\cite{AguilarSaavedra:2004wm}. With more data and further refined analysis methods some of these scenarios could be discovered or excluded, possibly already based on LHC Run-2 datasets.

\begin{figure}[tbp]
  \begin{center}
	\includegraphics[width=0.4\textwidth]{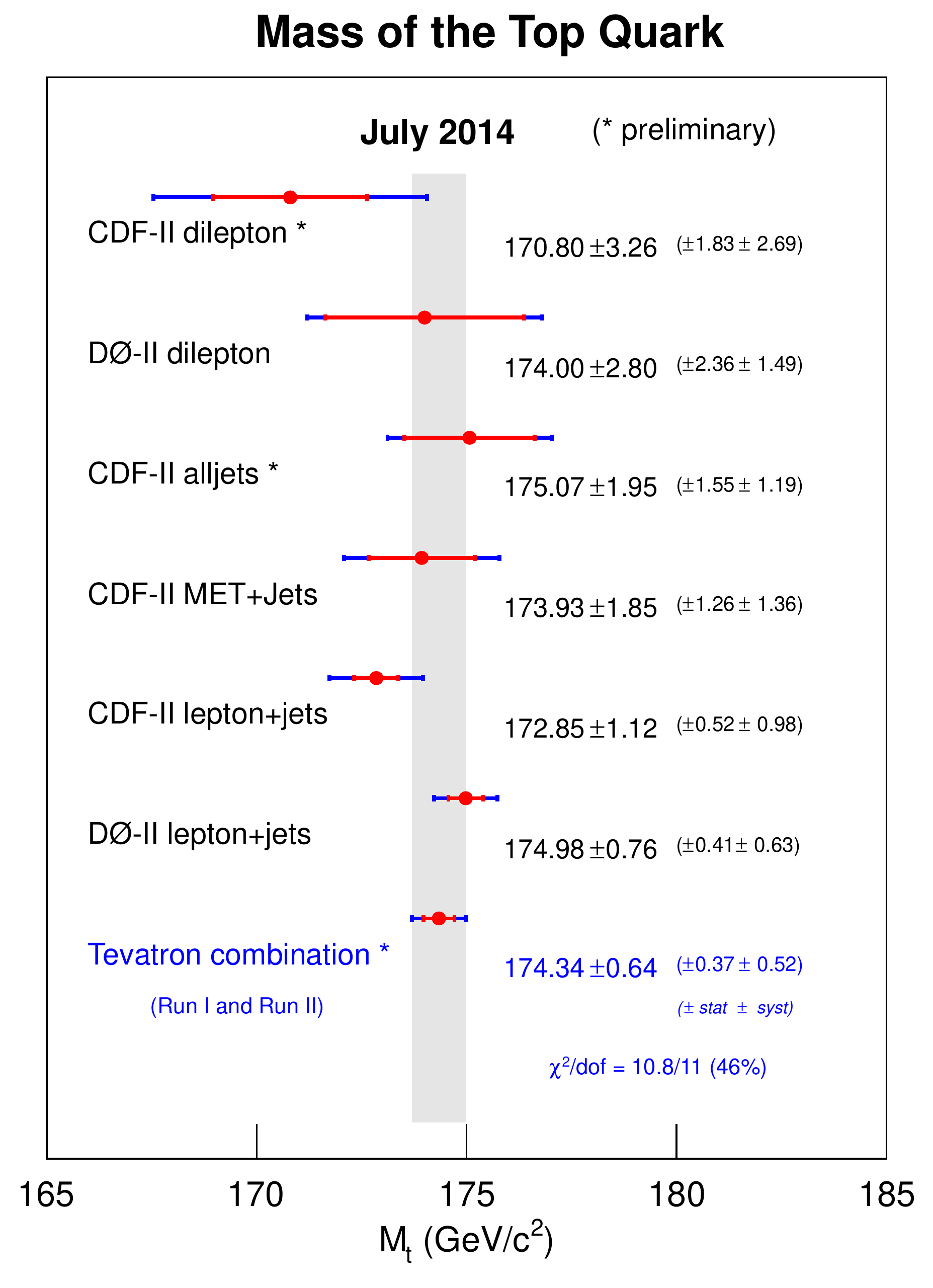}
	\includegraphics[width=0.57\textwidth]{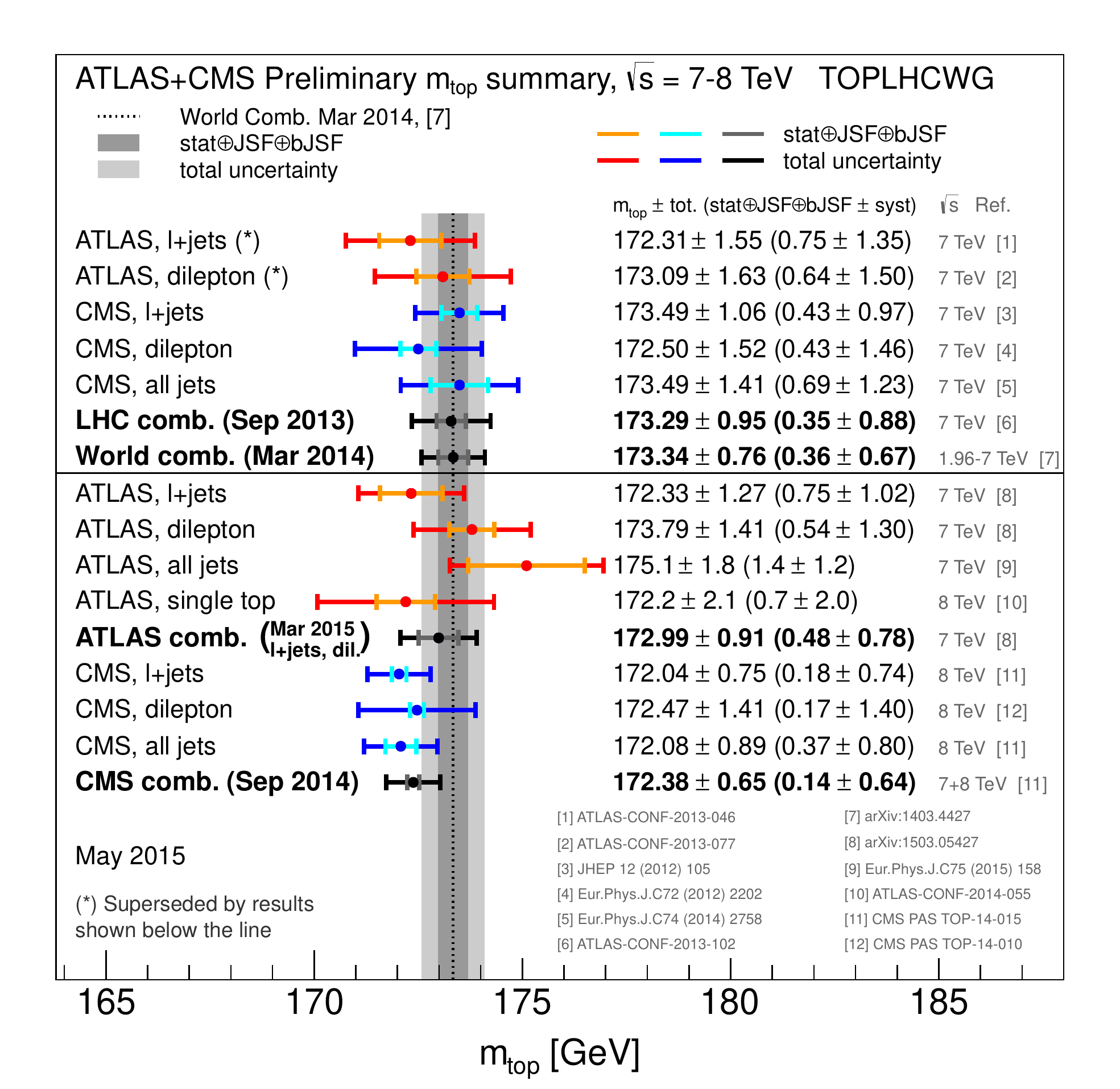}
	\caption{Summaries of mass measurements (left) from the Tevatron~\cite{Tevatron:2014cka} and (right) from the LHC~\cite{lhctopwg}.}
    \label{fig:top:mass:summary}
  \end{center}
\end{figure}

\section{Top-Quark Mass}

The mass of the top quark (or alternatively the Yukawa-coupling of the top quark to the Higgs boson) is a free parameter of the standard model. For a given top mass and the CKM matrix elements corresponding to the top quark, the standard model makes testable predictions for all top-quark properties. Conversely, precise property measurements provide for stringent consistency tests of the SM. Theoretically, the top-quark mass definition requires a renormalisation scheme. In the so-called pole mass scheme, the mass is defined as the pole in the renormalised quark propagator. This pole mass scheme is closely related to the intuitive understanding of the mass of a free particle. 

Experimentally, the mass of the top quark is conventionally determined by comparing suitable reconstructed distributions of the top-quark decay products in the data with those from simulation. The mass parameter in the simulation is then adjusted such as to optimally describe the data. These ``standard'' measurements of the top-quark mass have achieved a precision of better than 1 GeV. There is no well-defined relation of the mass parameters in simulations with a theoretically well-defined top-quark mass. However, quantitatively, the pole mass and the mass measured from final state reconstruction are expected to agree within $O(1~\rm GeV)$~\cite{Moch:2014tta}.

\subsection{Standard Top-Quark Mass Measurements}

\begin{figure}[tbp]
  \begin{center}
	\includegraphics[width=0.5\textwidth]{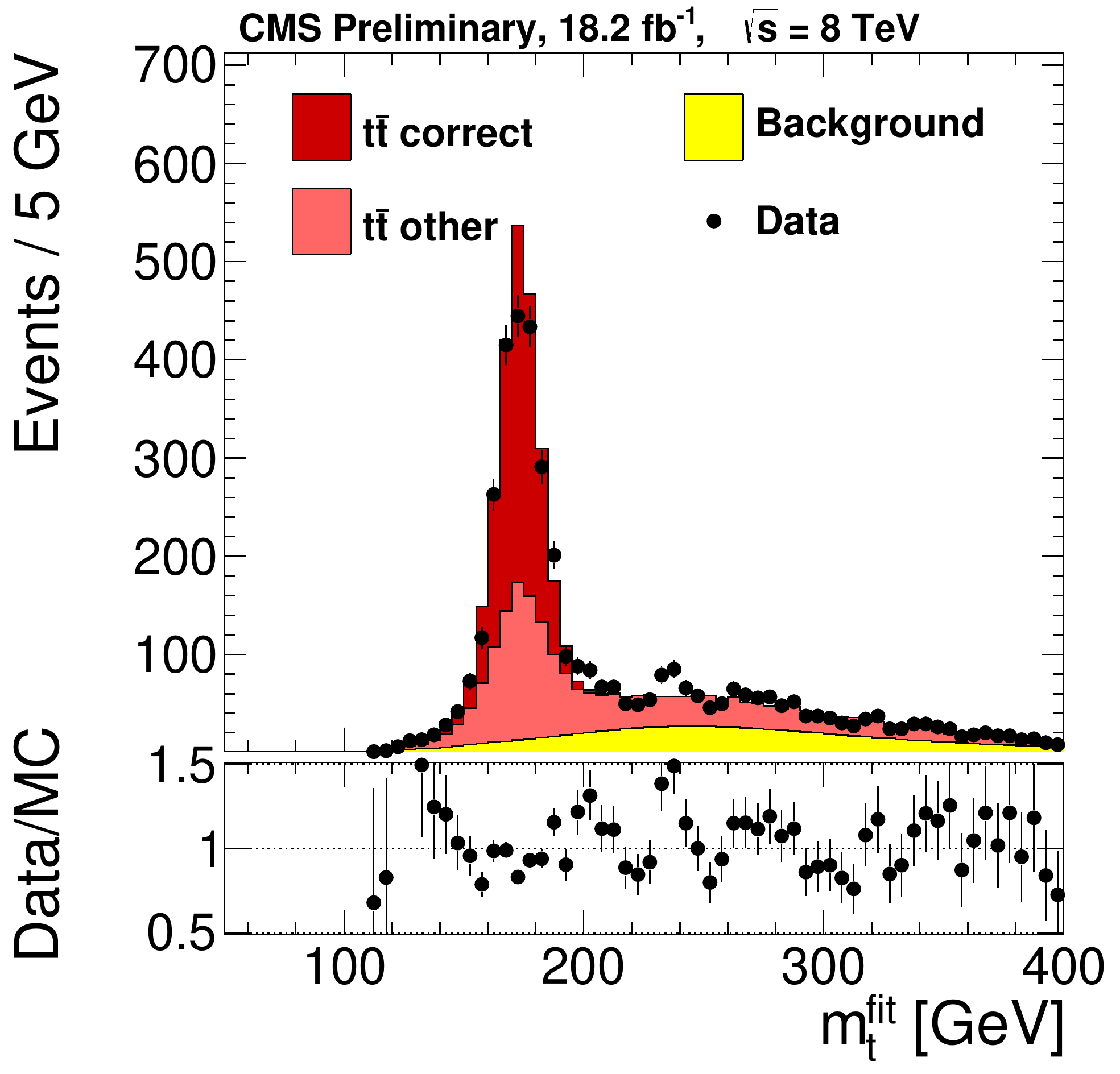}
	\includegraphics[width=0.4\textwidth]{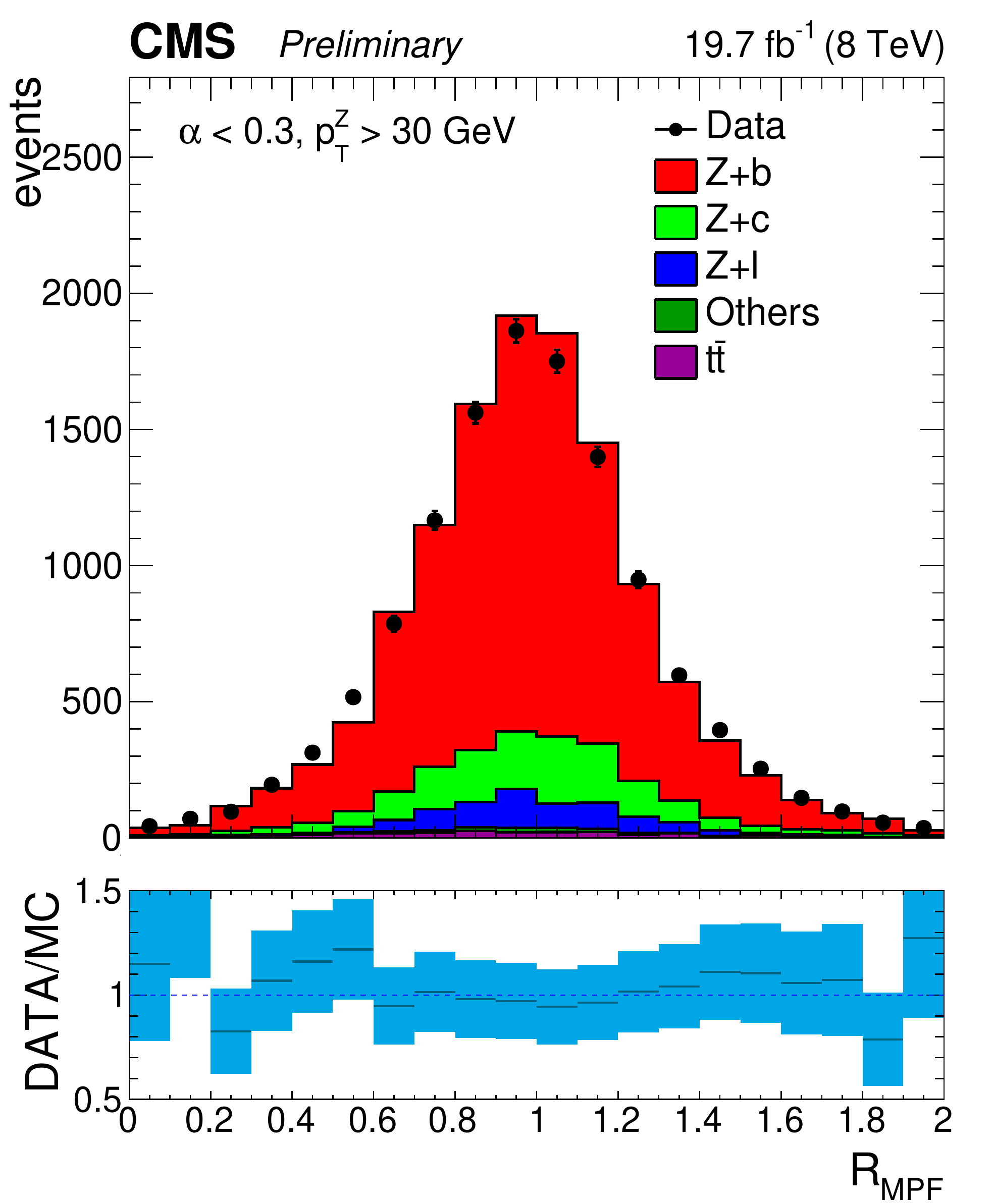}
	\caption{Left: distribution of the invariant mass of the top quark as obtained from a kinematic fit to the data. Also shown is the Monte Carlo simulation using a top-quark mass of 172.5 GeV. The multijet background is determined from the data~\cite{cms-top-14-002}. Right: balance distributions for b-jets relative to $Z$ boson transverse momenta~\cite{cms-jme-13-001}.}
    \label{fig:top:mass:cms}
  \end{center}
\end{figure}

Overviews of the results from standard top-quark mass measurements at the Tevatron and at the LHC are shown in Figure~\ref{fig:top:mass:summary}. In the following a few of the most recent measurements are described in more detail.

Using the full dataset at $\sqrt{s}=8$ TeV, the CMS experiment performed a top-quark mass measurement using $t\bar{t}$ events in which both top quarks decay hadronically~\cite{cms-top-14-002}. The presence of six jets is required of which two have to be identified as arising from b-quarks. A kinematic fit is used to assign the final-state jets to W bosons and top-quark candidates. 
In the fit the top-quark mass is determined simultaneously with an overall jet energy scale factor (JSF), constrained by the known mass of the W boson. Hypotheses are rejected if the goodness-of-fit probability is smaller than 20\%. In Figure~\ref{fig:top:mass:cms} (left) the distribution is shown. A clear narrow peak is seen on a relatively small background. The mass is determined in a joint maximum-likelihood fit to the selected events. Dominant uncertainties arise from the jet energy scale, the modelling of flavour-dependent jet energy corrections and the modelling of pile-up events. The top-quark mass is measured to be $172.08 \pm 0.36 (stat+JSF) \pm 0.83 (syst)$ GeV. 

Using the unprecedentedly large dataset at the LHC, CMS was able to perform a direct determination of the residual b-jet energy scale corrections from the data~\cite{cms-jme-13-001}. Events are considered in which a Z-boson decaying into a muon or electron pair is balanced in the transverse plane against a b-jet. An energy scale correction specific to b-jets is then estimated by comparing the jet energy distributions in data and Monte Carlo simulation. From the distribution shown in Figure~\ref{fig:top:mass:cms} (right) a value for the correction of $ 0.998 \pm 0.004(stat) \pm 0.004(syst) $ is obtained. Its consistency with unity leads to the conclusion that no additional b-jet-specific energy scale correction is needed. 

\begin{figure}[tbp]
  \begin{center}
	\includegraphics[width=0.48\textwidth]{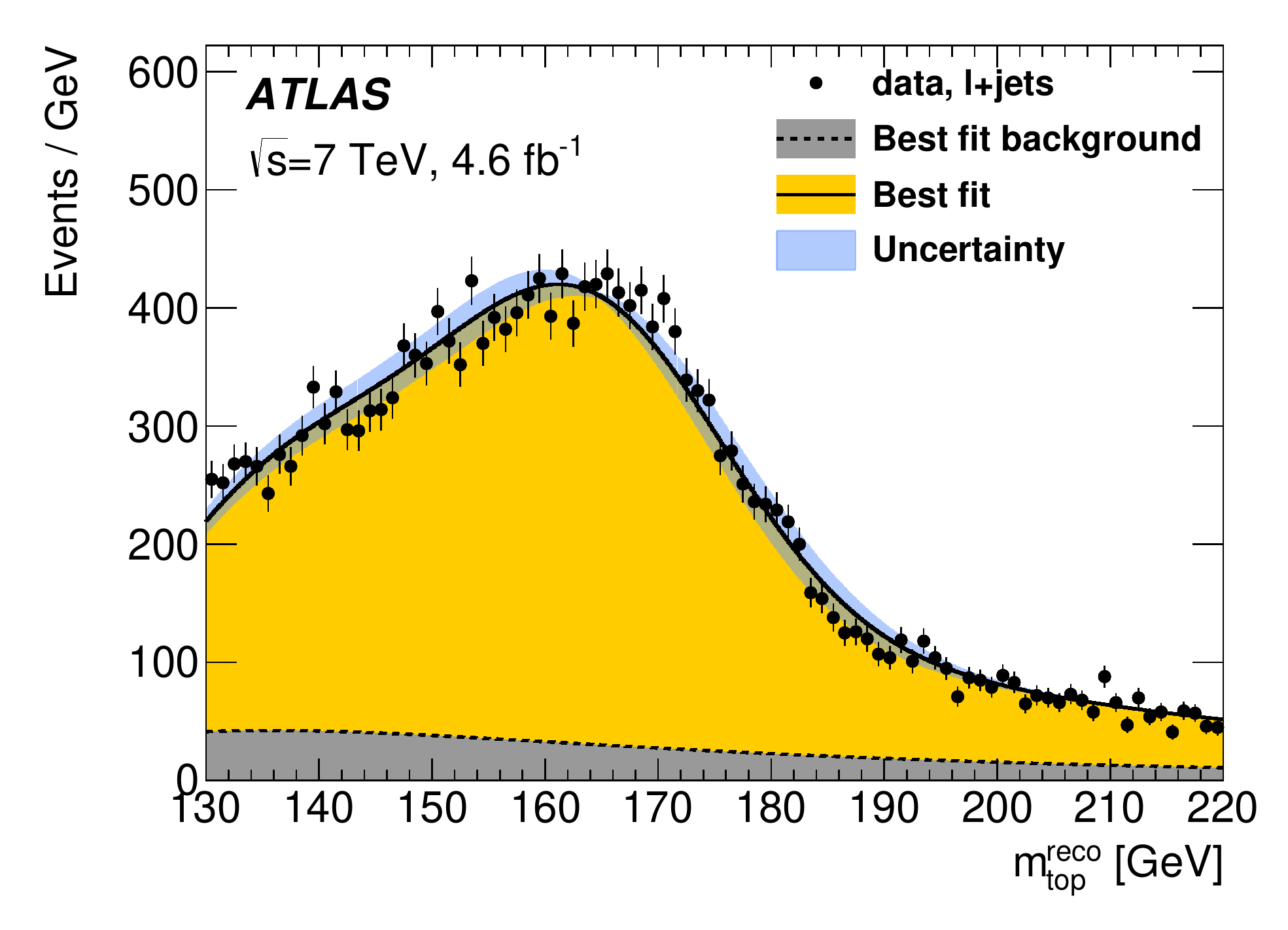}
	\includegraphics[width=0.48\textwidth]{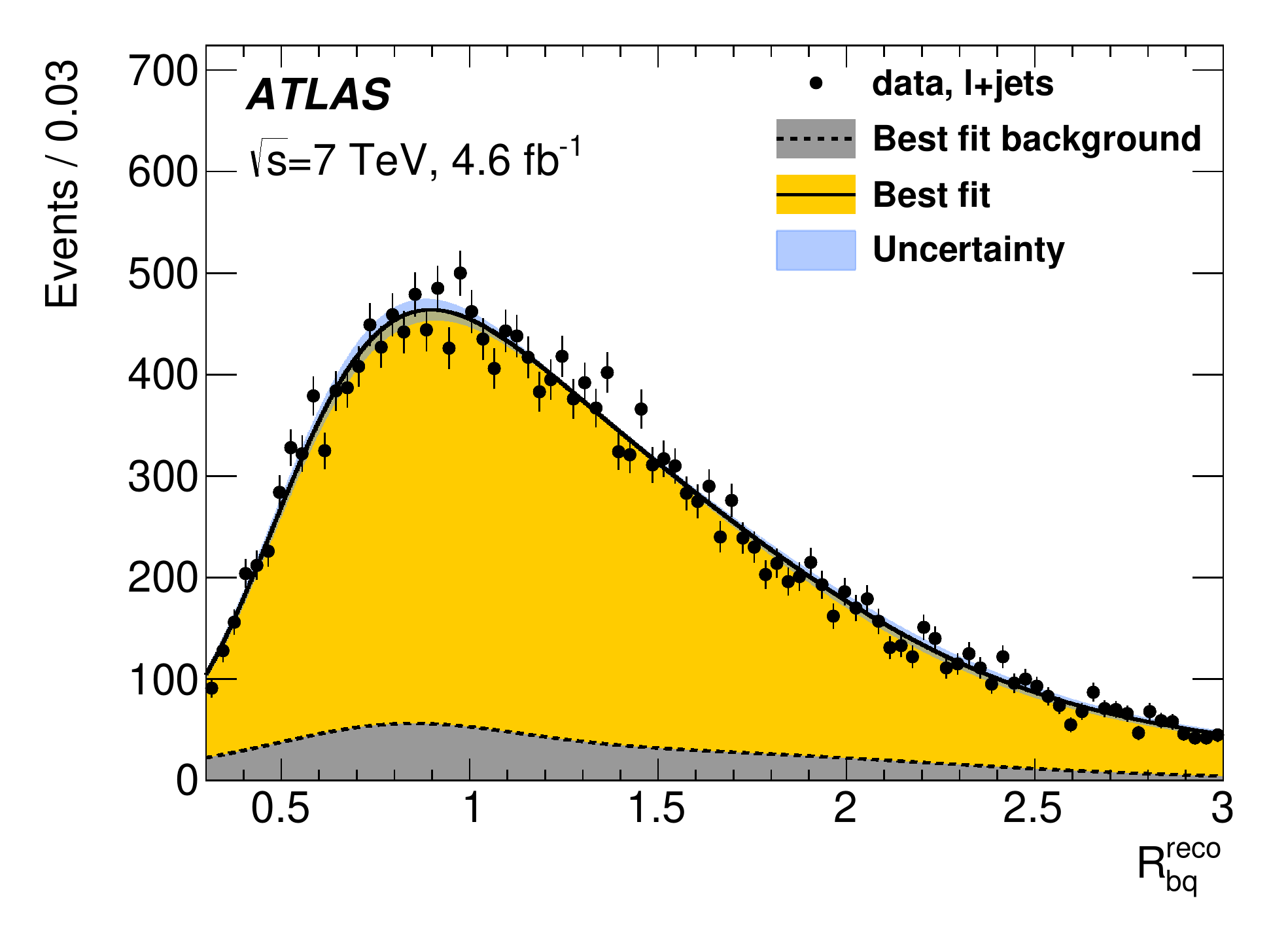}
	\caption{Reconstructed distributions of the top-quark mass and of the observable $R_{bq}$ which is used to constrain the b-jet specific energy scale (see text)~\cite{Aad:2015nba}.}
    \label{fig:top:mass:atlas}
  \end{center}
\end{figure}

The ATLAS experiment published a measurement of the top-quark mass from analyses of the dilepton and lepton+jets channels~\cite{Aad:2015nba}. A kinematic likelihood fit is used to reconstruct the event kinematics and to find the most likely assignment of reconstructed jets to partons. In the lepton+jets channel a three-dimensional template technique is used to determine the top-quark mass simultaneously with a correction for the global jet energy scale, using a constraint to the mass of the W boson, and an additional correction for the b-jet energy scale using the observable $R_{bq}$. The latter observable, derived as the ratio of the scalar sum of transverse momenta of b-tagged jets over the scalar sum of transverse momenta of the two jets associated with the hadronic W boson decay, is sensitive to the b-jet energy response, and independent of the top-quark mass. In Figure~\ref{fig:top:mass:atlas} the reconstructed distributions of data and simulation are displayed for the top-quark mass and the observable $R_{bq}$. The combination of the the 3d-result in the lepton+jets channel with the result in the dilepton channel yields a measured top-quark mass $m_{top}=172.99 \pm 0.48(stat) \pm 0.78(syst)$ GeV. Both the statistical and systematic uncertainties are expected to decrease further with more statistics.

The CDF and D0 experiments each report new top-quark mass measurements in the dilepton channel using the full Tevatron datasets. The CDF experiment uses a two-variable approach~\cite{Aaltonen:2015hta}. A kinematic reconstruction algorithm is used determine a preferred reconstructed top-quark mass for each event. This procedure makes use of the full event information. To minimise the impact of the jet energy scale uncertainty, a likelihood fit is performed including a second variable $M^{\rm alt}_{lb}$ which makes use of the energies and opening angles of the leptons with respect to the corresponding b-jets. The measurement yields a value for the top-quark mass of $171.5 \pm 1.9 (stat) \pm 2.5 (syst)$ GeV. 

The D0 experiment makes use of a neutrino weighting technique for the kinematic event reconstruction~\cite{D0:2015dxa}. To optimise statistical uncertainties the analysis makes use of the multiple kinematic solutions per event. The top-quark mass is determined from a likelihood fit of the Monte Carlo simulation for different values of $m_{top}$ to the data. The measurement yields $173.32 \pm 1.36(stat) \pm 0.85(syst)$ GeV.

\begin{figure}[tbp]
  \begin{center}
	\includegraphics[width=0.48\textwidth]{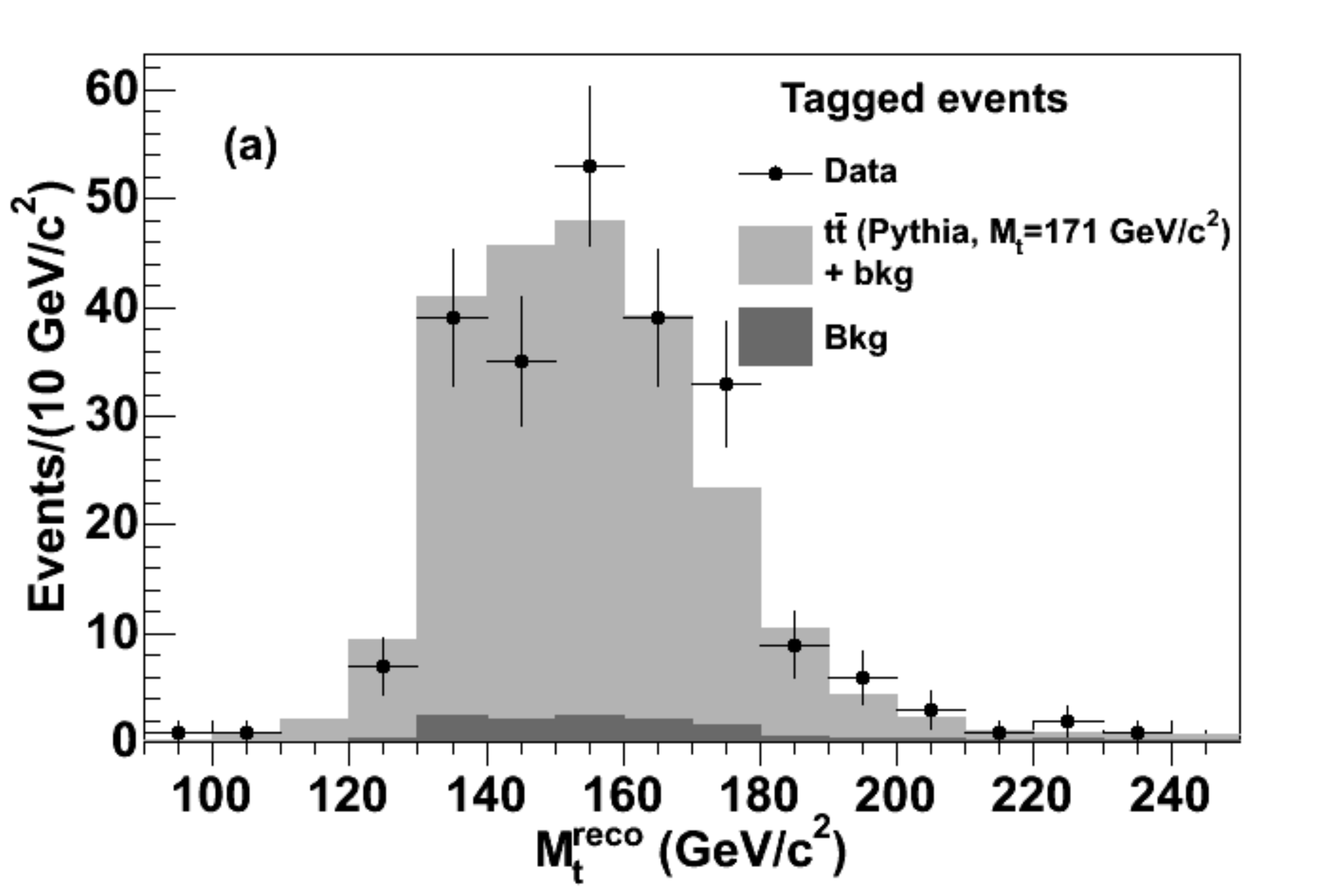}
	\includegraphics[width=0.42\textwidth]{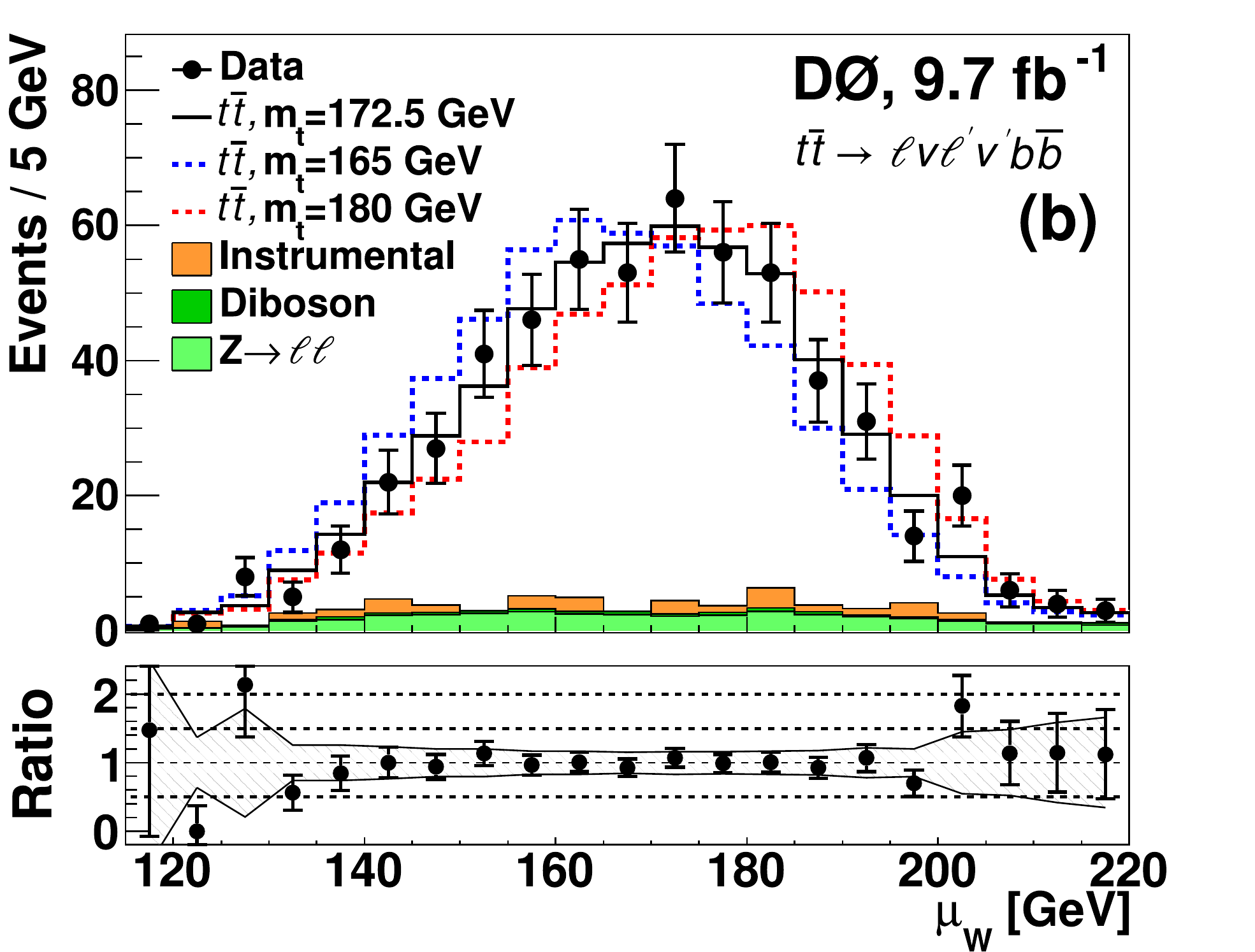}
	\caption{Left: distribution of reconstructed top-quark mass in the CDF experiment~\cite{Aaltonen:2015hta}. Right: distribution of the mass estimator for the final event sample in the D0 experiment~\cite{D0:2015dxa}.}
    \label{fig:top:mass:tevatron}
  \end{center}
\end{figure}

\subsection{Top-Quark Pole Mass Measurements}

\begin{figure}[tbp]
  \begin{center}
	\includegraphics[width=0.55\textwidth]{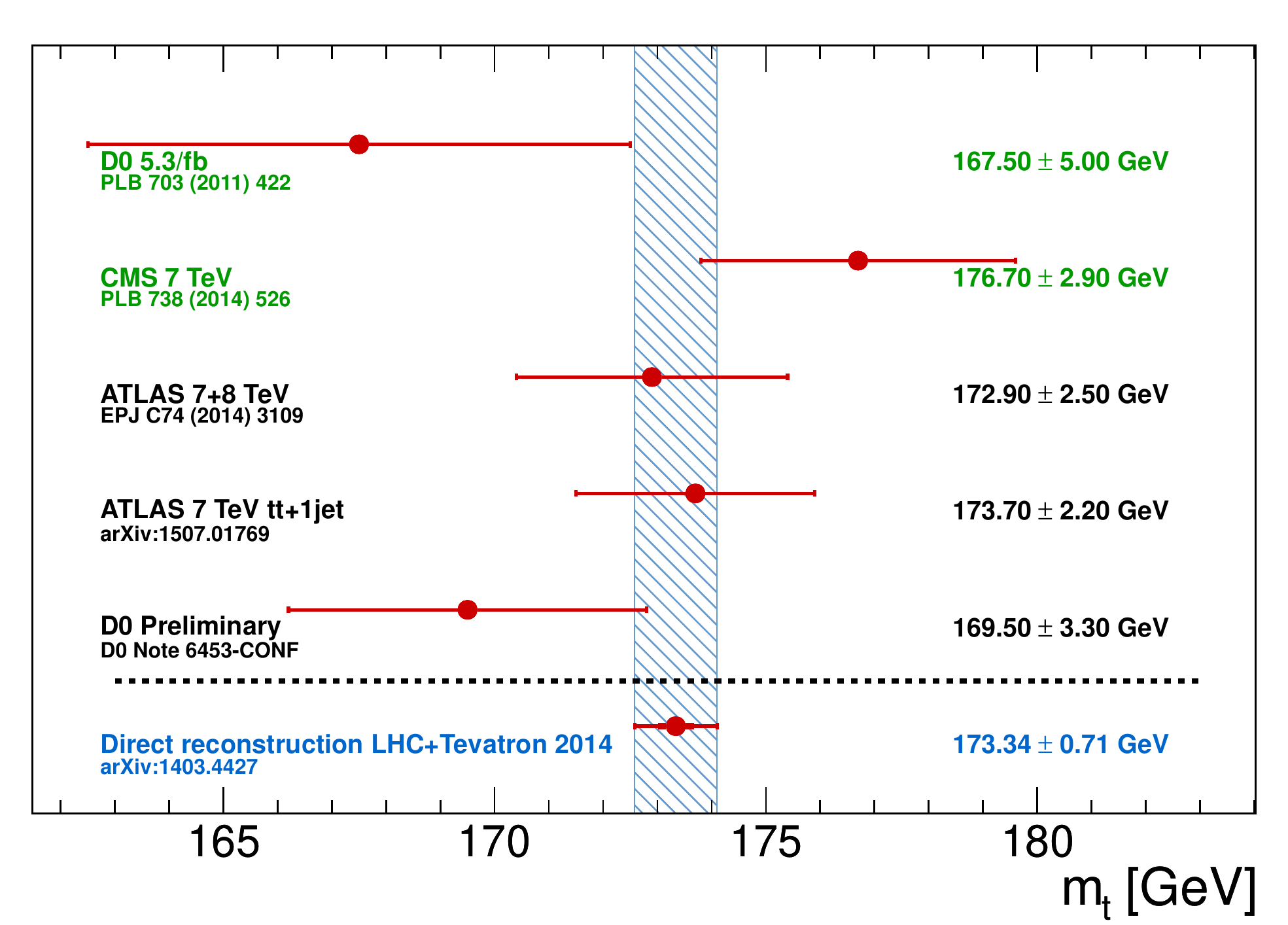}
	\includegraphics[width=0.43\textwidth]{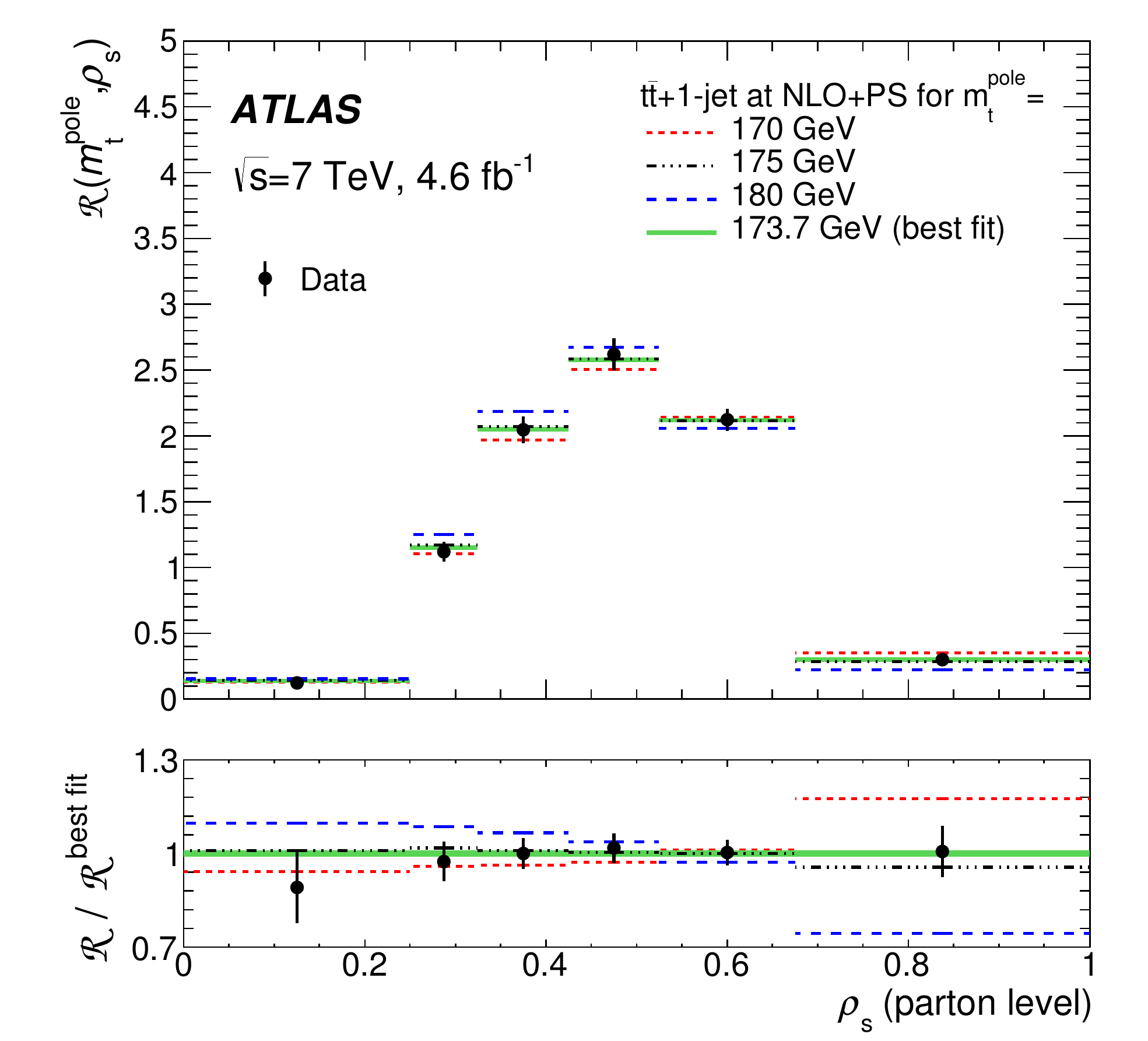}
	\caption{Left: summary of top-quark pole mass measurements. Right: distribution of the invariant mass of the $t\bar{t}+1$-jet system~\cite{Aad:2015waa}.}
    \label{fig:top:polemass}
  \end{center}
\end{figure}

An alternative approach to standard techniques for the measurement of the top-quark mass is to extract its value from the measured inclusive $t\bar{t}$ cross section. This approach has the advantage that the cross section and the pole mass are directly related, such that the extraction yields a theoretically well-defined quantity. The D0, CMS and ATLAS experiments have used their cross-section measurements to extract the top-quark pole mass~\cite{Chatrchyan:2013haa,Aad:2014kva,cms-top-13-004,D0-6453-CONF} as defined at NNLO accuracy~\cite{Czakon:2013goa}. The extractions are performed for different parton distribution functions and take into account the experimental dependence of the measured cross section on the assumed top-quark mass. A summary of the results is shown in Figure~\ref{fig:top:polemass}. 

The ATLAS experiment has presented a measurement of the top-quark pole mass using the differential $t\bar{t}$ cross section as a function of the invariant mass of the $t\bar{t}+1$-jet system~\cite{Aad:2015waa}. This distribution provides information of the top-quark mass via the mass-dependent threshold and cone-effects for the radiation of hard gluons~\cite{Alioli:2013mxa}. The ATLAS analysis is based on the dataset at a centre-of-mass energy of 7 TeV. The measured distribution is compared to the prediction at next-to-leading-order accuracy in quantum chromodynamics. The measured value of the top-quark pole mass is $173.7 \pm 1.5 (stat) \pm 1.4 (syst) ^{+1.0}_{-0.5} (theory)$ GeV. 

\section{Conclusions}

New precise results on top-quark properties continue to become available. The Tevatron experiments CDF and D0 and the LHC experiments ATLAS and CMS provide complementary information. The Tevatron measurements report Legacy results using the full Tevatron dataset, and the ATLAS and CMS experiments are concluding the LHC Run-1 analyses. A detailed picture of the top quark has been established, confirming the top-quark properties as compliant with standard model expectations. LHC Run-2 has started and until the year 2018, an integrated luminosity of about 100 fb$^{-1}$ at a centre-of-mass 13 TeV is expected, yielding one order of magnitude more top-quark events than collected so far. The increased statistics will give access to yet another new realm of precision. With the new data, and further progress in experimental and theoretical methods, measurements will reach unprecedented precision, and top quarks will become even more powerful probes for new physics searches.


\begin{thebibliography}{99}

\bibitem{cristinziani}
Markus Cristinziani, ``Top-Quark Production'', these proceedings.

\bibitem{atlas}
  G.~Aad {\it et al.} [ATLAS Collaboration],
  JINST {\bf 3} (2008) S08003.

\bibitem{cms}
  S.~Chatrchyan {\it et al.} [CMS Collaboration],
  JINST {\bf 3} (2008) S08004.

\bibitem{Aad:2014mfk}
G.~Aad {\it et al.} [ATLAS Collaboration],
Phys.\ Rev.\ Lett.\  {\bf 114} (2015) 14,  142001, [arXiv:1412.4742 [hep-ex]].

\bibitem{Chatrchyan:2013wua}
S.~Chatrchyan {\it et al.} [CMS Collaboration],
Phys.\ Rev.\ Lett.\  {\bf 112} (2014) 18,  182001, [arXiv:1311.3924 [hep-ex]].

\bibitem{cms-top-14-005} CMS Collaboration, CMS-PAS-TOP-14-005.

\bibitem{cms-top-13-015} CMS Collaboration, CMS-PAS-TOP-13-015.

\bibitem{acth} J.\,K\"uhn and G.\,Rodrigo, hep-ph/9802268, hep-ph/9807420.

\bibitem{Czakon:2014xsa}
  M.~Czakon, P.~Fiedler and A.~Mitov,
  Phys.\ Rev.\ Lett.\  {\bf 115} (2015) 5,  052001,
  [arXiv:1411.3007 [hep-ph]].

\bibitem{Aaltonen:2012it}
  T.~Aaltonen {\it et al.} [CDF Collaboration],
  Phys.\ Rev.\ D {\bf 87} (2013) 9,  092002,
  [arXiv:1211.1003 [hep-ex]].

\bibitem{Abazov:2014cca}
  V.~M.~Abazov {\it et al.} [D0 Collaboration],
  Phys.\ Rev.\ D {\bf 90} (2014) 072011,
  [arXiv:1405.0421 [hep-ex]].

\bibitem{melnikov}
Kiril Melnikov, ``Top-Quark Theory'', these proceedings.

\bibitem{Khachatryan:2015oga}
  V.~Khachatryan {\it et al.} [CMS Collaboration],
  arXiv:1507.03119 [hep-ex].

\bibitem{Khachatryan:2015mna}
  V.~Khachatryan {\it et al.} [CMS Collaboration],
  arXiv:1508.03862 [hep-ex].

\bibitem{lhctopwg}
  https://twiki.cern.ch/twiki/bin/view/LHCPhysics/LHCTopWGSummaryPlots 

\bibitem{Abazov:2015fna}
  V.~M.~Abazov {\it et al.} [D0 Collaboration],
  Phys.\ Rev.\ D {\bf 92} (2015) 5,  052007,
  [arXiv:1507.05666 [hep-ex]].



\bibitem{AguilarSaavedra:2004wm}
  J.~A.~Aguilar-Saavedra,
  Acta Phys.\ Polon.\ B {\bf 35} (2004) 2695,
  [hep-ph/0409342].

\bibitem{Aad:2015uza}
  G.~Aad {\it et al.} [ATLAS Collaboration],
  arXiv:1508.05796 [hep-ex].

\bibitem{cms-top-14-003} CMS Collaboration, CMS-PAS-TOP-14-003.

\bibitem{Aad:2015gea}
  G.~Aad {\it et al.} [ATLAS Collaboration],
  arXiv:1509.00294 [hep-ex].

\bibitem{cms-top-14-007} CMS Collaboration, CMS-PAS-TOP-14-007.

\bibitem{Aad:2014dya}
  G.~Aad {\it et al.} [ATLAS Collaboration],
  JHEP {\bf 1406} (2014) 008,
  [arXiv:1403.6293 [hep-ex]].

\bibitem{cms-top-14-019} CMS Collaboration, CMS-PAS-TOP-14-019.

\bibitem{cms-top-13-017} CMS Collaboration, CMS-PAS-TOP-13-017.


\bibitem{Moch:2014tta}
  S.~Moch {\it et al.},
  arXiv:1405.4781 [hep-ph].

\bibitem{Tevatron:2014cka}
  Tevatron Electroweak Working Group [CDF and D0 Collaborations],
  arXiv:1407.2682 [hep-ex].


\bibitem{cms-top-14-002} CMS Collaboration, CMS-PAS-TOP-14-002.

\bibitem{cms-jme-13-001} CMS Collaboration, CMS-PAS-JME-13-001.

\bibitem{Aad:2015nba}
  G.~Aad {\it et al.} [ATLAS Collaboration],
  Eur.\ Phys.\ J.\ C {\bf 75} (2015) 7,  330, 
  [arXiv:1503.05427 [hep-ex]].

\bibitem{Aaltonen:2015hta}
  T.~Aaltonen {\it et al.} [CDF Collaboration],
  Phys.\ Rev.\ D {\bf 92} (2015) 032003, 
  [arXiv:1505.00500 [hep-ex]].

\bibitem{D0:2015dxa}
  V.~M.~Abazov {\it et al.} [D0 Collaboration],
  Phys.\ Lett.\ B {\bf 752} (2016) 18,
  arXiv:1508.03322 [hep-ex].

\bibitem{Chatrchyan:2013haa}
  S.~Chatrchyan {\it et al.} [CMS Collaboration],
  Phys.\ Lett.\ B {\bf 728} (2014) 496
   [Phys.\ Lett.\ B {\bf 728} (2014) 526],
  [arXiv:1307.1907 [hep-ex]].

\bibitem{Aad:2014kva}
  G.~Aad {\it et al.} [ATLAS Collaboration],
  Eur.\ Phys.\ J.\ C {\bf 74} (2014) 10,  3109,
  [arXiv:1406.5375 [hep-ex]].

\bibitem{cms-top-13-004} CMS Collaboration, CMS-PAS-TOP-13-004.

\bibitem{D0-6453-CONF} D0 Collaboration, D0 Note 6453-CONF.

\bibitem{Czakon:2013goa}
  M.~Czakon, P.~Fiedler and A.~Mitov,
  Phys.\ Rev.\ Lett.\  {\bf 110} (2013) 252004,
  [arXiv:1303.6254 [hep-ph]].

\bibitem{Aad:2015waa}
  G.~Aad {\it et al.} [ATLAS Collaboration],
  JHEP {\bf 1510} (2015) 121,
  [arXiv:1507.01769 [hep-ex]].

\bibitem{Alioli:2013mxa}
  S.~Alioli {\it et al.},
  Eur.\ Phys.\ J.\ C {\bf 73} (2013) 2438,
  [arXiv:1303.6415 [hep-ph]].



\end{thebibliography}
\end{document}